# APPLICATION OF THE QUADRATURES METHOD TO THE NLS WITH SATURATION

D. S. Levko

Institute of Physics NAS Ukraine

**E-mail:** d.levko@gmail.com

**Abstract.** The solution of nonlinear Schrödinger equation with saturation was found by means the quadratures method in terms of degeneracy theory. It was shown the existence conditions for soliton solutions.

**1. Introduction**

At the present time the solitons theory are penetrated at different branches of modern physics: plasma physics, nonlinear optics, hydrodynamics, the condensed matter physics and others. Every branch has its own soliton's equation. It is Korteveg - De Vries equation (KdV) [1], sin-Gordon equation (sG) [2], or Nonlinear Schrödinger equation (NLS) [3]. Also, it is existed generalizing equations such as mixed KdV (also named as Gardner's equation) [4], Gross-Pitaevskii equation [5].

Though, NLS has a special role among the named equations. It consists with the next reasons [6]. At first, the function in equation is complex. At second, the dispersion of NLS is the same as in the case of free relativistic particle. The NLS of special kind are found the application at the present time. Such equations are solved in nonlinear optics for description of the propagation of the surface solitons [7].

It is existed many exact methods for solving the NLS. It is the inverse method, Riemann's method, method of finite zone integration. But special NLS doesn't have such solutions. For this goal we must to use alternative methods. One of them is quadratures method [8]. In present work this method was used for NLS with saturation.

**2. Description of the model**

The NLS with non-Kerr nonlinearity is nonintegrable and doesn't allow the soliton solutions in general case. It consists with the distortion of balance between the nonlinear processes and dispersion or diffraction.

At the present time the NLS with saturation are found many applications in connection with problems in nonlinear optics [7]:

$$iq_t + \frac{1}{2}q_{xx} + \frac{q|q|^2}{1+S|q|^2} = 0. \tag{1}$$

Here $q(x,t)$ is the complex function, $S$ is the saturation parameter. The inverse problem is not known for (1). As a result, we need other method to solve it. In this article the quadratures method was used [8]. It doesn't allow to recognize the solitons and solitary waves because it doesn't allow to explore the waves' collisions. Though, its using allows to find the solutions of NLS-likes equations.

Soliton-like equations contain one parameter family of solutions with changed amplitude in Hamiltonian systems with conservation of energy

$$q(x,t) = f(x) \cdot \exp(i\omega t + i\varphi). \tag{2}$$

Here $f(x)$ is a real periodical function, $\varphi$ is a phase which depends from spatial variable $x$ in general case: $\varphi = \varphi(x)$. The form (2) is analogous to [8] with one difference: now the phase time is spatial coordinate $x$.

Let us find the stationary solution. So, the phase is constant and the (1) is

$$\frac{1}{2}f'' - \omega f + \frac{f^3}{1+Sf^2} = 0. \tag{3}$$

Now function $f$ is the position of particle moving in the field with potential

$$U(f) = \int_0^{f^2} \frac{f^2}{1+Sf^2} df^2 - \omega f^2 = \left(\frac{1}{S} - \omega\right) \cdot f^2 - \frac{1}{S^2}\ln(1+Sf^2). \tag{4}$$

The equation of motion has a form

$$f_{xx} = -\frac{\partial U}{\partial f}. \tag{5}$$

It is the problem to find the exact quadrature in the right side of equality:

$$x - x_0 = \pm \int \frac{df}{\sqrt{2(E - U(f))}}. \tag{6}$$

### 3. Numerical solution

The potential (4) and the phase portraits with different energies in (6) are presented on fig. 1. Equation (3) contains soliton solutions only for phase trajectory with $E = 0$ (so called separatrix). The motion occurs between the stop point (it is the point not achieved through the finite time) and the turning point (it is the point achieved through the finite time). The first coincides with the origin of coordinates. In case $E > 0$ we get one oscillating solution and in case $E < 0$ the degeneracy has place: we get two oscillating solutions.

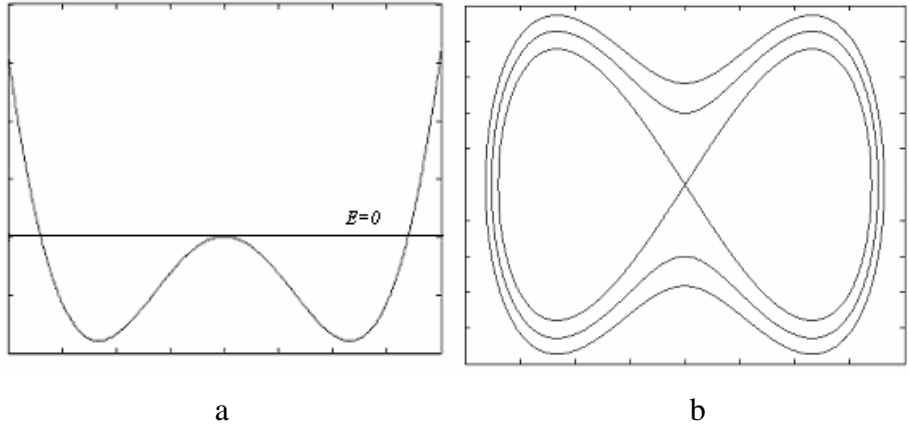

a  b

Figure 1. a) effective potential, b) phase portraits

The variable $I = f^2$ is the intensity of the light field in the problems of nonlinear optics. In term of this quantity the potential (4) can be write in the form:

$$U(I) = \left(\frac{1}{S} - \omega\right) \cdot I - \frac{1}{S^2} \ln(1 + SI). \qquad (7)$$

It allows directly to get the conditions of the soliton solutions' existence. The intensity increases from zero. So, the potential has to reach the minimum and after that it has to growth to zero value. It is achieved for $U(I) > 0$. Because of this, the result condition is:

$$\omega < \frac{1}{S}. \qquad (8)$$

In the case of small $S\omega$ we can find the maximum of intensity:

$$I_m \approx 2\omega\left(1 + \frac{4}{3}S\omega\right). \qquad (9)$$

The results of numerical solutions of (3) are presented on fig. 2-4. It is seen that on separatrix we get a soliton solution. The bold curve is the right branch of potential solution. In the case $E > 0$ the solution is two periodical functions that shifted in space. It is consisted with shift of the initial conditions. The case with $E < 0$ is the case of degeneracy with solutions like two symmetrical harmonic function. The both of them are the right and left branches of potential well. Its depth defines the solution's amplitude.

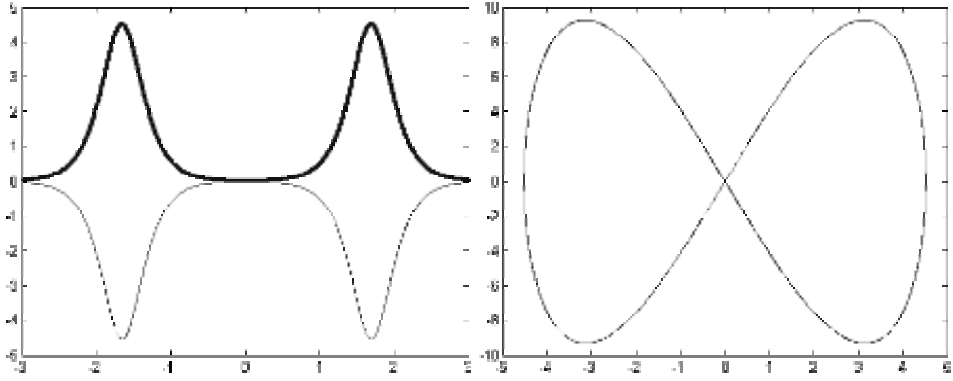

Figure 2. Solutions of (3) and corresponding phase portrait for $E = 0$

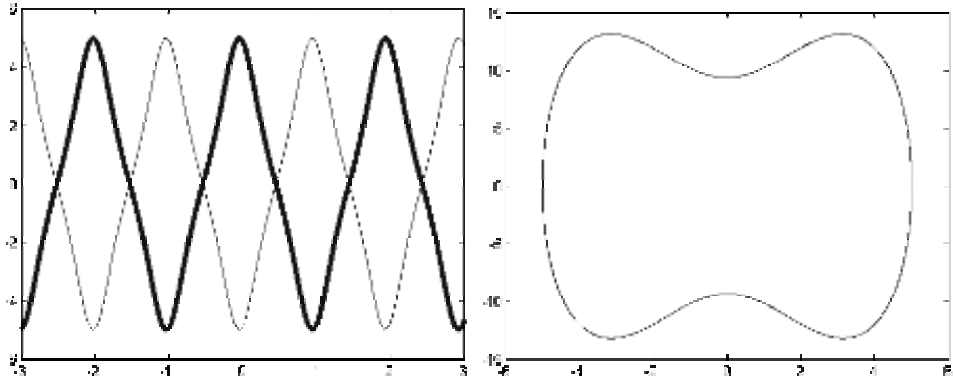

Figure 3. Solutions of (3) and corresponding phase portrait for $E > 0$

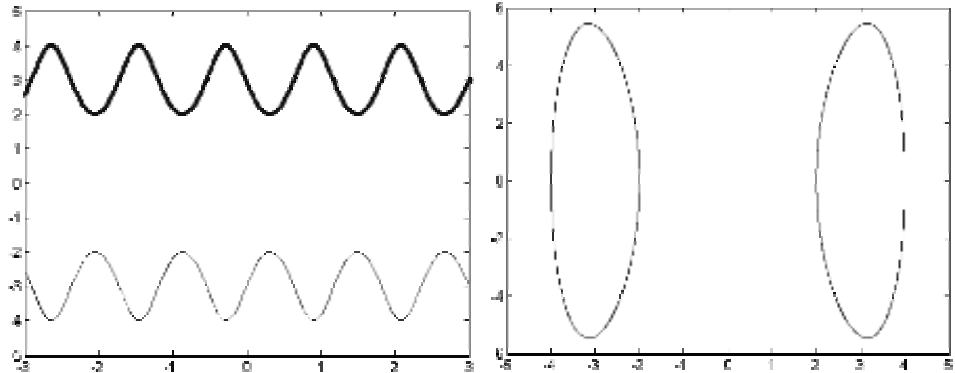

Figure 4. Solutions of (3) and corresponding phase portrait for $E < 0$

**4. Analytical solutions**

Let us find the solutions of equation (3). We can develop the logarithm in (7) in set in the case of small amplitude of saturation parameter $S$:

$$\ln(1 + Sf^2) \approx Sf^2 - \frac{S^2 f^4}{2} + \frac{S^3 f^4}{3} + \dots . \tag{10}$$

At first, let us limit by two first terms (it will lead to disappearing of the saturation effects) and consider the motion of the particle on separatrix. Than (1) is simplified to general NLS with known solution

$$f(x) = 2\sqrt{2\omega} \cdot \text{sech}\sqrt{2\omega}x. \tag{11}$$

We get two solutions on the curvatures with nonzero energy $E \neq 0$

$$f_{1,2}(x) = \frac{\sqrt{\alpha_{1,2}}}{\text{sn}(\sqrt{2E\alpha_{1,2}}\,x)}, \tag{12}$$

where $\alpha_{1,2} = -\frac{\omega}{2E} \pm \frac{1}{2E}\sqrt{\omega^2 + 2E}$. The absolute value of elliptical integral is $k_{1,2}^2 = \frac{\alpha_{2,1}}{\alpha_{1,2}}$.

If we take into account the third term in developing of potential we get the next separatrix solution

$$f(x) = \frac{\sqrt{f_2}}{\sqrt{1 - \frac{f_2}{f_1} \cdot \text{th}^2 \frac{x\varphi}{2}}} \cdot \text{sech}\frac{x\varphi}{2}, \tag{13}$$

$$f_{1,2} = \frac{3}{4S}\left(1 \mp \sqrt{1 - \frac{16S\omega}{3}}\right), \quad \varphi = \frac{\sqrt{2\omega} \cdot \left(1 - \sqrt{1 - \frac{16S\omega}{3}}\right)}{\sqrt{1 - \frac{16S\omega}{3}}}. \tag{14}$$

It is the soliton solution because with infinite phase times it is decreasing with the exponential law. The transferred energy by such object is

$$E_{sol} = \int_{-\infty}^{+\infty} |f|^2 dx = -\frac{4f_1}{\varphi} \cdot \ln\left|\frac{\sqrt{f_1} - \sqrt{f_2}}{\sqrt{f_1} + \sqrt{f_2}}\right|. \tag{15}$$

The point with zero coordinate and momentum is hyperbolical and the motion here is unstable (fig. 1). Though, the solution on zero separatrix has soliton's properties.

The next step is the setting of (4) near the extremums $f_{a,b} = \pm\sqrt{\frac{\omega}{1-S\omega}}$. If we consider only quadratic terms we get two shifted oscillating solution with distance between $2f_a$

$$f(x) = f_{a,b} + \sqrt{\frac{E-b}{a}} \cdot \sin(\sqrt{2a}\,x). \tag{16}$$

Here two quantities were introduced

$$b = U(f_{a,b}) = \frac{\omega}{S} - \frac{1}{S^2}\ln(1 - S\omega), \quad 2a = U''(f_{a,b}) = \frac{4(1-S\omega)^2}{S}. \tag{17}$$

The solution (16) is conforming the earliest statement: the amplitude of solution defines by the well's depth.

It is seen from (17) that constant $a$ is positive. Also, it is seen from (8) that $b$ is negative. In addition, the last parameter is the minimum of potential energy. Because of this $E-b$ is positive ($E<0$, $|E|<|b|$) and the function (16) is real with any meanings of parameters.

Separatrix solution is a soliton's couple if we take into account four first terms in setting of potential. It is the right and the left solitons (fig. 2)

$$f(x) = \pm \frac{2\sqrt{a} \cdot \exp(\sqrt{a}x)}{1 + b \cdot \exp(2\sqrt{a}x)}. \tag{18}$$

Here $a = 2\omega$, $b = \frac{3}{2} - 2S\omega$. Since the both parameters are always positive the function $f(x)$ is always real.

With the named reasons the presence of two potential's (4) minimums defines the degree of degeneracy, the position of extremum (stopping point). The number of minimums defines the quantity of solitons. Also, the depth of potential well defines the amplitude of oscillating solutions. In connection with this we can change (4) by means of the next function

$$U_{eff}(f) = a \cdot f^4 - b \cdot f^2. \tag{19}$$

In this case we can get the exact solution. In (19) $a$ and $b$ are some positive constants. Its fashion will lead to the equality of zero's positions and extremums of (4) with the named points for modeling potential (19).

## 5. Conclusion

The solution of NLS with saturation parameter was considered in the present work. The branches of existence of soliton and oscillating solutions were found in terms of quadratures method. Also the results of numerical modeling were presented. The comparison showed the good agreement between the both.

The results of analytical solution of the stationary equation in terms of degeneracy theory were presented. It is also in good agreement with the numerical modeling.

**References**

[1] C. Gardner et al. 1967 Method for solving the Korteveg-de Vries equation *Phys. Rev. Lett.* **19** 1095-97

[2] M.J. Ablowitz et al. 1973 Method for solving the Sine-Gordon equation *Phys. Rev. Lett.* **30** 1262-64

[3] L. Di Menza, C. Gallo 2007 The black solitons of one-dimensional NLS equations *Nonlinearity* **20** 461-496

[4] Abdul-Majid Wazwaz 2007 New solitons and kink solutions for the Gardner equation *Communications in Nonlinear Science and Numerical Simulation* **12** 1395-1404

[5] L.P. Pitaevskii 2006 Bose-Einstein condensates in a laser field *Uspekhi Fizicheskikh Nauk* **176** 345-364 (in Russian)

[6] V.G. Makhan'kov et al. 1994 Localized non-topological structures: construction of solutions and stability problems *Uspekhi Fizicheskikh Nauk* **164** 121-148 (in Russian)

[7] Ya. Kartashov et al. 2006 Surface vortex solitons *Preprint* physics/0601029

[8] A.G. Kornienko et al. 1997 Classification of the solutions of Nonlinear equations of Schrödinger's type of special kind *Moscow Univ. Bull.* **5** 10-13 (in Russian)